\documentclass[twocolumn,secnumarabic,amssymb, nobibnotes, aps, pra, superscriptaddress]{revtex4-2}
\usepackage{lipsum}
\usepackage{xcolor}
\usepackage{changes}
\usepackage{graphicx}
\usepackage{siunitx}
\usepackage{tikz}

\usepackage{ulem}
\usepackage{mathtools}
\usepackage{hyperref}
\hypersetup{
    colorlinks=true,
    linkcolor=blue,
    filecolor=blue,
    urlcolor=blue,
    citecolor=blue
    }
\usepackage[capitalise]{cleveref}

\newcommand*\circled[1]{\tikz[baseline=(char.base)]{
            \node[shape=circle,draw,inner sep=0.5pt] (char) {#1};}}

\begin{document}

\title{Deterministic preparation of a dual-species two-ion crystal\\
}

\author{Maximilian J. \surname{Zawierucha}}
\email[(he/him/his) ]{maximilian.zawierucha@ptb.de}
\thanks{Equal contribution}
\affiliation{Physikalisch-Technische Bundesanstalt, Bundesallee 100, 38116 Braunschweig, Germany}
\affiliation{Institut für Quantenoptik, Leibniz Universität Hannover, Welfengarten 1, 30167 Hannover}

\author{Till \surname{Rehmert}}
\thanks{Equal contribution}
\affiliation{Physikalisch-Technische Bundesanstalt, Bundesallee 100, 38116 Braunschweig, Germany}
\affiliation{Institut für Quantenoptik, Leibniz Universität Hannover, Welfengarten 1, 30167 Hannover}

\author{Jonas \surname{Keller}}
\affiliation{Physikalisch-Technische Bundesanstalt, Bundesallee 100, 38116 Braunschweig, Germany}

\author{Tanja E. \surname{Mehlstäubler}}
\affiliation{Physikalisch-Technische Bundesanstalt, Bundesallee 100, 38116 Braunschweig, Germany}
\affiliation{Institut für Quantenoptik, Leibniz Universität Hannover, Welfengarten 1, 30167 Hannover}

\author{Piet O. \surname{Schmidt}}
\affiliation{Physikalisch-Technische Bundesanstalt, Bundesallee 100, 38116 Braunschweig, Germany}
\affiliation{Institut für Quantenoptik, Leibniz Universität Hannover, Welfengarten 1, 30167 Hannover}
\author{Fabian \surname{Wolf}}
\email[(he/him/his) ]{fabian.wolf@ptb.de}
\affiliation{Physikalisch-Technische Bundesanstalt, Bundesallee 100, 38116 Braunschweig, Germany}

\date{\today}
\begin{abstract}

The demand for efficient preparation methods for dual-species ion crystals is rapidly expanding across quantum technology and fundamental physics applications with trapped ions. We present a deterministic and efficient technique to produce such crystals, utilizing the segmented structure of a linear Paul trap. By precisely tailoring the trapping potentials, we can split, move, and discard parts of an ion chain. This process is automated in a sequence that converts a larger ion sample into the desired configuration. A critical component of our approach is the accurate identification of crystal constituents. This is achieved by matching the measured positions of fluorescing ions against theoretical expectations for larger crystals, thus facilitating the detection of non-fluorescing ions and enabling accurate ion counting. We demonstrate that our method reliably produces two-ion crystals within minutes. These results represent a significant advance in the production of two-species ion crystals with applications ranging from quantum logic spectroscopy and optical clocks to quantum computing and simulation with trapped ions.\\
\end{abstract}
\maketitle

\section{Introduction}

Trapped ions have been established as a versatile platform for advancing the field of atomic, molecular, and optical physics over the past few decades. Unprecedented accuracies in trapped ion clocks \cite{ludlow_optical_2015, brewer__2019, huntemann_single-ion_2016, zhiqiang_176lu_2023, huang_liquid-nitrogen-cooled_2022}, highly sensitive sensors \cite{brownnutt_ion-trap_2015, wolf_motional_2019, campbell_rotation_2017, baumgart_ultrasensitive_2016, gilmore_quantum-enhanced_2021, biercuk_ultrasensitive_2010, mccormick_quantum-enhanced_2019} and quantum computers \cite{cirac_quantum_1995, blatt_entangled_2008, bruzewicz_trapped-ion_2019} are just some of the outstanding applications which are being pursued.
While many of these experiments are possible with a single species of ions, extending the control capabilities to dual-species ion systems enables a richer variety of experiments and is even a necessity for certain applications.
For instance, co-trapped ions of a different species can serve as an \textit{in situ} sensor for perturbing fields \cite{barrett_polarizability_2019, wolf_scheme_2023, steinel_evaluation_2023}, or for sympathetic cooling of an ion species with no (accessible) laser-cooling transition \cite{wubbena_sympathetic_2012, king_algorithmic_2021, rugango_sympathetic_2015, wan_efficient_2015,  sels_doppler_2022, meiners_towards_2018, cui_sympathetic_2018, calvin_rovibronic_2018, ohtsubo_frequency_2017, chen_sympathetic_2017, guggemos_sympathetic_2015, willitsch_chemical_2008, kwapien_sympathetic_2007, ryjkov_sympathetic_2006, ostendorf_sympathetic_2006, roth_sympathetic_2005}. The latter has enabled the development of optical quantum logic clocks \cite{schmidt_spectroscopy_2005, brewer__2019, cui_evaluation_2022, guggemos_frequency_2019, kramer_aluminum_2023}, precision spectroscopy on highly charged \cite{micke_coherent_2020, king_optical_2022} and molecular ions \cite{chou_frequency-comb_2020, wolf_non-destructive_2016} and helps to preserve coherence for quantum information processing \cite{wang_single-qubit_2017, wang_single_2021}.

Here, we present a technique for preparing such a dual-species two-ion crystal. The process relies on modifying the confining dc potentials to iteratively split a dual-species ion chain. The constituents of each sub-crystal are detected and the more favorable one is kept in the trap, while the other one is expelled from the trap.
Repeating this process successively reduces the size of the ion chain, finally leading to a dual-species two-ion crystal.
While the composition of the sub-samples after the splitting operation is probabilistic, the possibility of re-merging and repeating the splitting renders the entire preparation scheme deterministic. Ion detection is based on position-resolved fluorescence detection of one ion species and from that inferring the other non-fluorescing ion species' position.

Different techniques can be used to load ions into a radiofrequency trap, such as ionizing an atomic vapor from an electrically \cite{ballance_short_2017} or optically heated oven \cite{gao_optically_2021}, a laser cooled atomic cloud \cite{sage_loading_2012, bruzewicz_scalable_2016} or by ablating atoms with a laser \cite{Leibrandt_ablation_2007, wu_adaptively_2021}. All of those approaches have advantages and drawbacks.

Typically, oven loading is slower than ablation loading.
More rapid buildup of atom flux for faster loading can be achieved by increasing the oven temperature \cite{ballance_short_2018}, but leads to a high background pressure inside the vacuum chamber.
Loading from a pre-cooled ensemble of atoms requires substantial technical overhead and suitable laser cooling transitions for the atomic species.
For ablation loading, there is a trade-off between fast loading and control over the exact amount of ions, since the loading process is governed by Poisson statistics \cite{leibrandt_laser_2007}, which means that higher laser intensities yield a higher probability of successful loading, but results in a loss of control over the ion number. Additionally, variations in beam pointing, target surface degradation and laser intensity have a strong impact on the number of loaded ions \cite{shi_ablation_2023, white_isotope-selective_2022, wu_adaptively_2021, osada_deterministic_2022, vrijsen_efficient_2019, olmschenk_laser_2017, zimmermann_laser_2012}.

These limitations for ablation loading can be alleviated by subtractive preparation schemes, where a larger ion crystal is loaded and then reduced to the desired size by removing selected ions. In addition to the improved robustness of these schemes, large crystals of laser-cooled ions provide faster sympathetic cooling for co-trapped species that cannot be laser-cooled directly. This circumvents the problem of long crystalization times that have been observed for example in aluminum quantum logic clocks \cite{guggemos_sympathetic_2015, sun_efficiently_2022, kramer_aluminum_2023} or recapture of highly charged ions \cite{schmoeger_coulumb_2015}.

In previously demonstrated subtractive schemes \cite{fulford_radiofrequency_1980, londry_mass_2003, schmidt_mass-selective_2020}, the removed ions are selected by their charge-to-mass ratio making it difficult to control the exact number of remaining ions. Furthermore, different ionic species with similar charge-to-mass ratio cannot be individually removed, which can be relevant for molecular or highly charged ion experiments.

Unlike previously demonstrated methods, the ejection approach presented here is independent of the ion species being removed. As a result, this approach offers the capability to generate dual-species ion crystals with arbitrary composition. An extension to multi-species ion crystals is possible if the composition of the ensemble can be determined.

\section{Experimental setup}

\begin{figure}
\begin{minipage}{2.4cm}
\includegraphics[width=\textwidth]{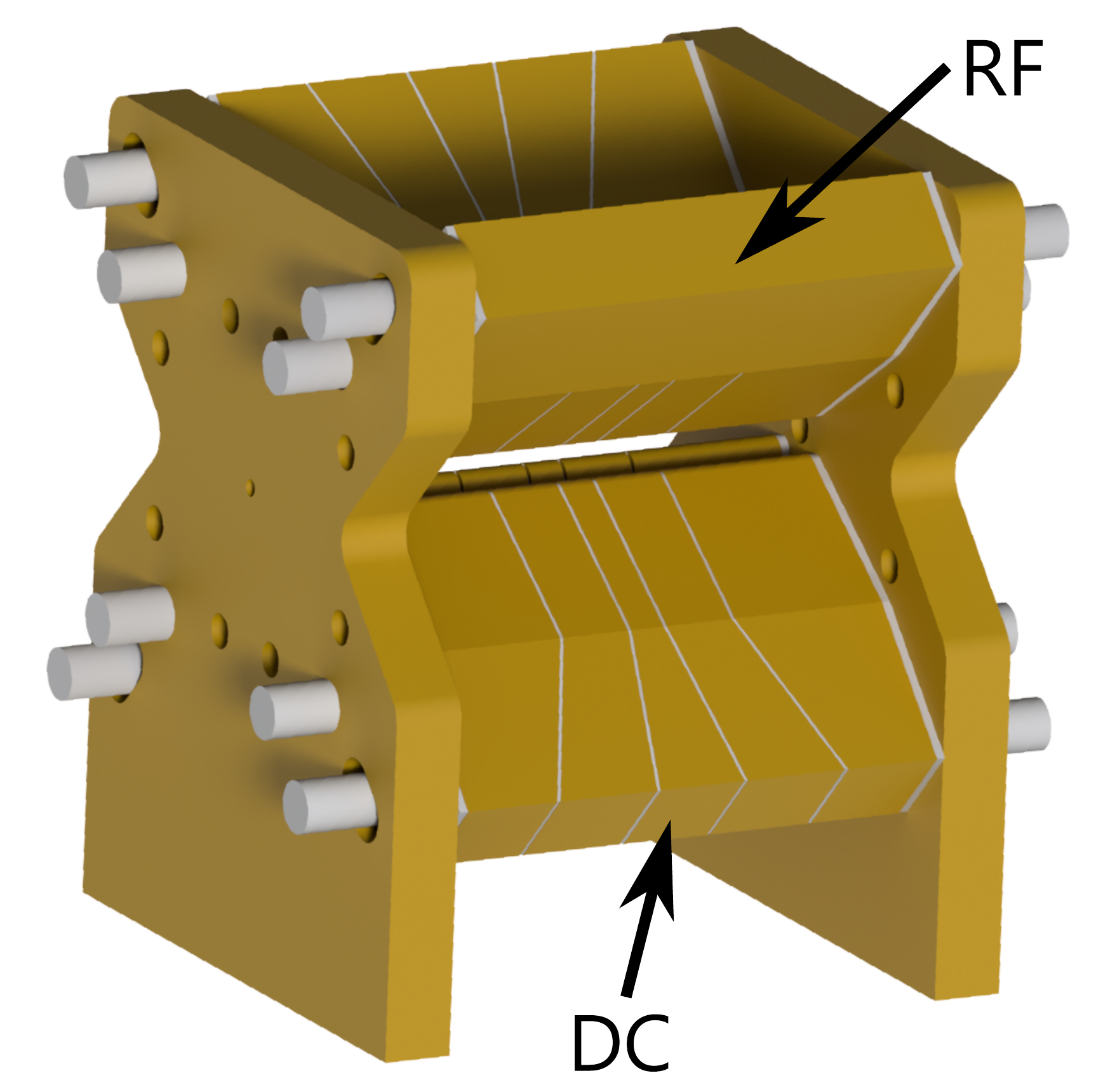} 
\label{trap-CAD}
\end{minipage}%
\begin{minipage}{6.2cm}
\includegraphics[width=\textwidth]{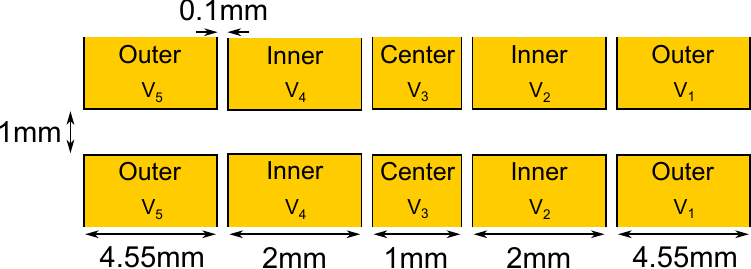}
\label{trapnomenclature}
\end{minipage}
\caption{Ion trap geometry. Left: CAD render of the ion trap used in the described experiments.
It is constructed using gold coated alumina pieces. Segmentation of dc electrodes is achieved by laser cutting.
The top right and bottom left blades carry rf signal at \SI{23.456}{\MHz} for radial confinement of the ions. The other two blades carry adjustable dc electric potentials for axial confinement of the ions and for performing ion-chain shuttling and splitting operations.
Right: Nomenclature and sizes of the dc segments.
}
\label{trap-cad-nomenclature}
\end{figure}

The experiment at hand is a trapped ion spectroscopy setup.
We use a linear Paul trap, for trapping ions at ultrahigh vacuum (\SI{<e-10}{\milli\bar}).
A CAD-render of the trap can be found in \cref{trap-cad-nomenclature}. The distance between electrodes and the ions is \SI{0.5}{\mm}.
The two opposing dc blades are segmented into five electrodes each, which can be individually addressed to provide axial confinement. With these segments, a double-well potential, resulting in two axially separated trapping regions, can be realized. A more detailed description of a similar trap and its construction can be found in \cite{leopold_cryogenic_2019}.
The experiments described here were performed with $^{40}\text{Ca}^+$ and $^{25}\text{Mg}^+$ ions. The calcium ions are laser-cooled with light at \SI{397}{\nm}. The resulting fluorescence of the ions is imaged on an Electron Multiplying Charge-Coupled Device (EMCCD) camera using a single aspheric lens. The magnesium ions are not laser-cooled and therefore not directly detectable by the imaging system, but appear as dark defects in the coulomb crystal.

Both ion species are loaded via ablation loading with a matchbox-sized laser at \SI{515}{\nm} (Mountain Photonics GmbH, IOP 0515L-21C-NI-NT-NF). The laser is pulsed with a repetition rate of \SI{3.5}{\kHz} and provides pulse powers of approximately \SI{30}{\micro\J}. It can either be focused onto a calcium or magnesium target inside the vacuum chamber, using a motorized mirror.
The ablated calcium ions are photo-ionized, resembling the scheme described in \cite{lucas_isotope-selective_2004}. 
Magnesium-25 is loaded from an enriched target without an ionization laser, similar to experiments with other species \cite{zimmermann_laser_2012}.

After ion loading, the trapping parameters fulfill  $\omega_{x,y} \gg \omega_z$, where $\omega_{x,y}$ are the radial and $\omega_z$ the axial trap frequencies of a single trapped ion. In our case these values are $\omega_x=2\pi \times$\SI{750}{\kHz}, $\omega_y=2\pi \times$\SI{950}{\kHz} and $\omega_z=2\pi \times$\SI{143}{\kHz}. This configuration results in the ions aligning in a linear chain along the axial trap direction, see \cref{ion-detection-pic}.

\section{Ion detection}
\label{Ion detection}
To determine the composition of a dual-species crystal containing calcium and magnesium ions, two detection steps are required.
For laser-cooled calcium ions the employed imaging system resolves the individual ions, which are typically separated by a few micrometer for the used trap settings, see \cref{ion-detection-pic}~\textbf{a)}. Therefore the calcium ions' positions can be extracted by detecting fluorescence peaks in the camera image.
These measured positions are compared to pre-calculated  theoretical values \cite{james_quantum_1998} for ion crystals of different sizes, revealing information about the number of non-fluorescing magnesium ions present in the crystal. 
This process is discussed in detail in the following sections.

\subsection{Bright ion detection}
To detect the bright calcium ions' positions and quantity, an image is taken using an  EMCCD camera with a \SI{14}{\bit} brightness resolution and an exposure time of \SI{0.2}{\s}, see \cref{ion-detection-pic}~\textbf{a)}. The image quality is compromised by noise, predominantly electronic noise, background photons and defective (bright) pixels.
In order to mitigate this noise, the image is post-processed.
False ion detections due to single bright pixels are effectively suppressed by applying a gaussian filter with a width of one pixel and setting pixels below a certain threshold to zero brightness. A good initial guess for this threshold is provided by the "otsu" threshold determination from the python  scikit-image module (skimage.filters.threshold\_otsu, V0.20.0, \cite{scikit-image}).
Lower detection errors are obtained by increasing this threshold by 200 (out of the full dynamic range of \SI{14}{\bit}), which was determined in a heuristic approach.
A detailed description of the Otsu threshold determination can be found in reference \cite{otsu_threshold_1979}. 

    Afterwards, the peak detection function (skimage.feature.peak\_local\_max, \cite{scikit-image}) from scikit-image is used to determine the bright ions' positions with a resolution of one pixel. To avoid double counting of single ions, we chose a minimum distance of six pixels (approx. \SI{4}{\um}) between detected ion positions.

The bright ion detection works well for linear ion chains and slight zig-zag configurations with up to approximately $20$ ions.
An example for this process can be found in \cref{ion-detection-pic}~\textbf{b)}.

\subsection{Dark ion detection}

In the experiments described here, magnesium ions are not laser-cooled and thereby cannot be detected via fluorescence imaging. However, when the bright calcium ions' positions are determined, the dark magnesium ions quantity and location in the crystal can be inferred. This is done by comparing pre-calculated equilibrium positions to the detected positions of the calcium ions.
A similar method has been used to count ions in a long chain, where only the central part of the chain was imaged \cite{kamsap_experimental_2016}.

Since the ion chain is mostly aligned along the vertical camera axis, we only consider the ions' vertical position component.

The equilibrium positions of ions in a linear chain are calculated following \cite{james_quantum_1998}.
In order to map the calculated ion positions to an expected pixel position on the camera $P^\text{theo}_j$, a calibration is needed, as described in 
            Appendix \ref{sec:app-cam-calib}.

The total number of bright and dark ions $n$ is determined by minimizing the measure $b(n)$, defined as
\begin{align}
b(n) = \sum_{i}\frac{\min_{\forall j}\left| P_i-P^\text{theo}_j(n) \right|^2}{n_{\text{bright}}},
\end{align}

which is the mean distance of the measured ion positions $P_i$ of the fluorescing ions to the closest theoretically determined position $P^\text{theo}_j(n)$. $n_\text{bright}$ is the number of fluorescing ions, determined by the previously described measurement.
The number of dark ions $n_\text{dark}$ is determined by subtracting $n_\text{bright}$ from $n$.
For the minimization of $b(n)$, the total number of ions was limited to $n<\min(2n_\text{bright},23)$. This choice helps to mitigate detection errors as described in Section~\ref{benchmark}.
The normalization of $b(n)$ to the number of bright ions is analogous to a reduced $\chi^2$ analysis and allows comparing the quality of the fit between crystals of different sizes.

\begin{figure}
\includegraphics[width=8.6cm]{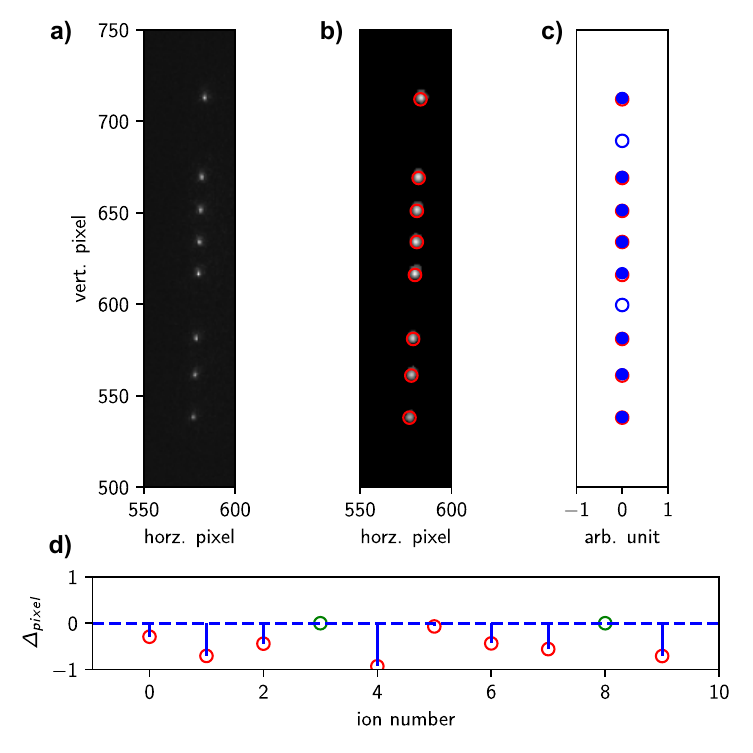}
\caption{
Step by step illustration of the ion chain detection process.
\textbf{a$)$} Unprocessed image of the ion chain taken with the EMCCD camera
\textbf{b$)$} Applied gauss and otsu filters and a peak detector to find the ion positions. Red circles visualize bright calcium ions detected by the algorithm.
\textbf{c$)$} Result of the dark ion detection. Empty circles indicate dark ion positions. The eight previously detected bright ions are indicated by filled circles.
\textbf{d$)$} Deviation of the detected ion positions from the best matching calculated ion chain.
}
\label{ion-detection-pic}
\end{figure}

The result of such a dark ion position determination can be found in \cref{ion-detection-pic}~\textbf{c)}, with typical deviation between detected and calculated ion positions below one pixel, see \cref{ion-detection-pic}~\textbf{d)}.

\subsection{Benchmark and limitation}\label{benchmark}
\begin{figure*}
\centering
\includegraphics[width=17.2cm]{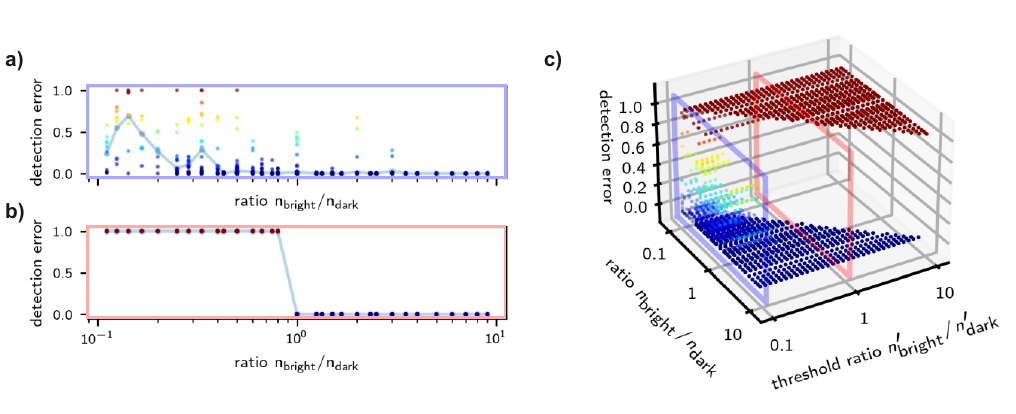}
\caption{Simulated detection error for ion crystals with $n\leq 10$ for different ratios between dark and bright ions. \textbf{a)} shows the detection error, where arbitrary ion configurations with $n'\leq 20$, where compared to the detected bright ion positions. Each scatter corresponds to one ion chain configuration. The blue line denotes the average detection error for a given bright to dark ion ratio. In subfigure~\textbf{b)} the detection error is shown with the additional restriction for the reference positions that more bright ions than dark ions are in the ion chain. The general case for different threshold ratios between dark and bright ions in the reference configurations is shown in subfigure~\textbf{c)}. See text for additional information.}
\label{fig:detection_error}
\end{figure*}

In order to benchmark the performance of the dark ion detection scheme, it was run on a dataset with sample size $205\,456$. The data was generated by recording ion chains with lengths of up to ten fluorescing ions.
The previously described process was used to infer the camera position $P_i$ of the fluorescing ions. Dual-species ion chains were simulated by selectively removing position data from the purely bright dataset. The dark ion detection algorithm was used to predict the ion chain configuration. In this implementation of the detection algorithm a maximum number of $20$ ions was assumed. \cref{fig:detection_error}~\textbf{a)} illustrates the detection error for different ratios of dark and bright ions. Each scatter represents one particular configuration of ions. Configurations with a larger fraction of dark to bright ions are more prone to detection errors. Also configurations with $n_\mathrm{bright}>n_\mathrm{dark}$ can be detected incorrectly. These false detections are caused by falsely assuming an ion chain with $n_\mathrm{bright}<n_\mathrm{dark}$. The resulting errors can be mitigated by restricting the theoretical configurations, that are compared to the bright ion positions to configurations with $n'_\mathrm{bright} \ge n'_\mathrm{dark}$. When describing restrictions for the detection algorithm, a prime $(')$ is added to the ion number.
\cref{fig:detection_error} shows that this restriction suppresses errors entirely for configurations with $n_\mathrm{bright}\ge n_\mathrm{dark}$, at the cost of restricting the detection algorithm to this subset of configurations. Experimentally, this precondition can be easily met by an appropriate choice of loading parameters. The condition $n'_\mathrm{bright} \ge n'_\mathrm{dark}$ can be generalized by introducing the limiting ratio $n'_\mathrm{bright}/n'_\mathrm{dark}$. The previously described precondition corresponds to $n'_\mathrm{bright}/n'_\mathrm{dark} \ge 1$. \cref{fig:detection_error}~\textbf{c)} shows the detection error in dependence of the limiting ratio and illustrates the trade-off between a low average detection error and a large number of detectable configurations. In all experiments described here, a limiting ratio of $n'_\mathrm{bright}/n'_\mathrm{dark}=1$ was used.
More details regarding the origin of the detection errors can be found in Appendix \ref{app-detection-errors}.

\

Additionally there are technical limitations of the detection scheme, restricting its application to $23$ ions with the parameters used.
Larger ion numbers lead to formation of 3D crystals for the chosen trapping parameters. These structures cannot be interpreted by our detection method.
To align larger ion crystals linearly, a higher radial confinement is necessary, which is infeasible in our setup.
In general, increasing the ion number increases the ion density in the center of the trap for a given axial confinement. Therefore it becomes more difficult to distinguish between ion chains with similar ion numbers.
Overcoming this limitation would require higher spatial resolution of the imaging system or operating at a lower axial confinement to get a larger separation of the ions.
The implementation presented here does only determine the ions position $P_i$ with one pixel resolution. More advanced peak detection methods could be used to interpolate the region between camera pixels, thereby providing higher resolution.
Decreasing the axial potential has its limits, as a very low axial confinement leads to fluctuating ion positions, because the position becomes much more sensitive to external electric field perturbations. Also, increasingly larger ion chains may no longer be within the detection region.

\section{Ion operations}
\label{Ion operations}

\begin{figure}
\includegraphics[width=8.6cm]{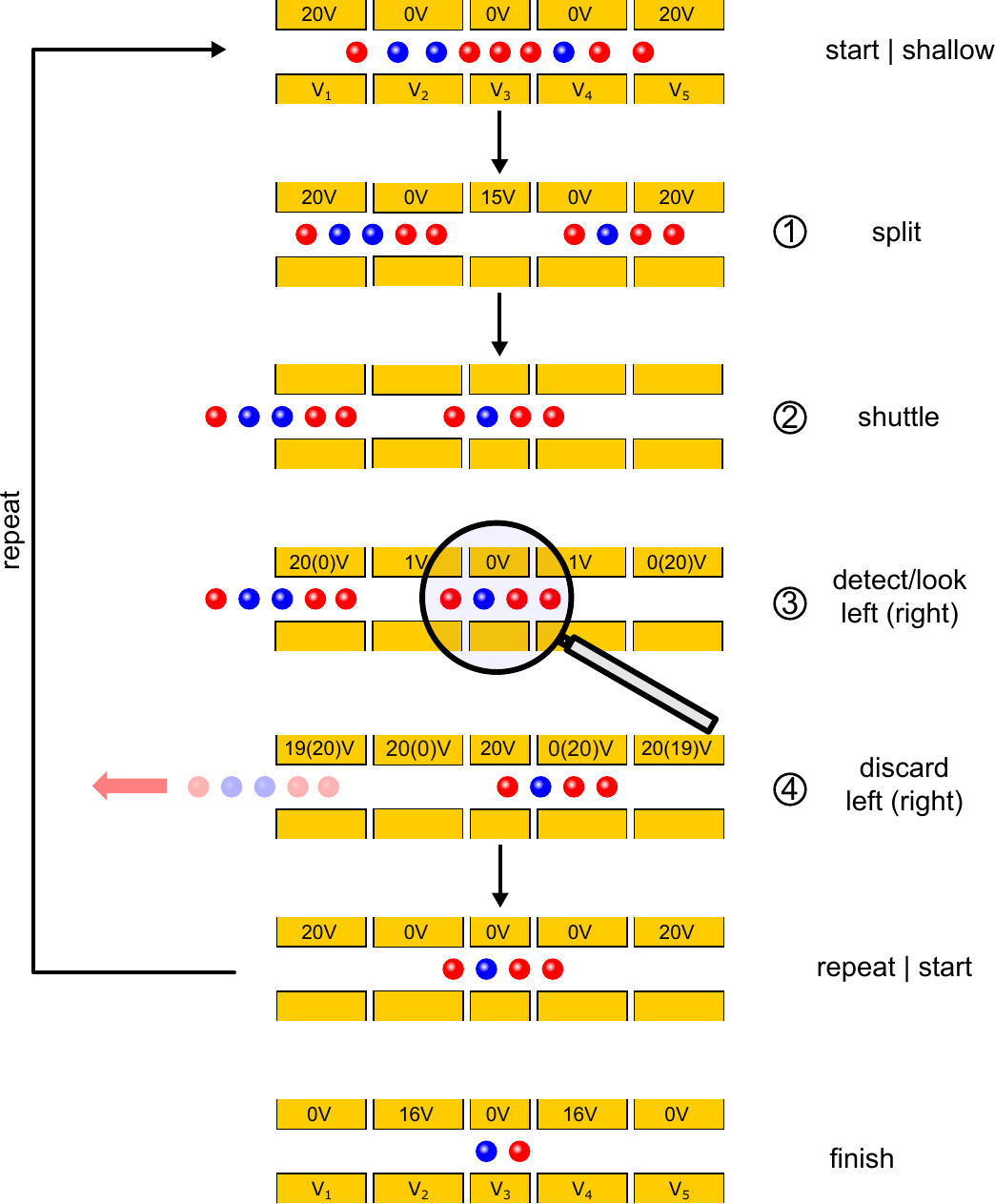}
\caption{Visualization of automated two ion crystal production loop. Blue and red balls depict the different ion species. Note: the ion position and distance with respect to the depicted segments is not to scale.}
\label{flow-chart}
\end{figure}

The presented ion crystal preparation procedure relies on splitting an ion chain into two parts, and discarding the side which is further from a desired target crystal. The decision logic is depicted in \cref{table-decision}.
In total, four different operations on the ion crystal are required: \circled{1} splitting the ion chain into two sub-chains, \circled{2} shuttling a sub-chain into the detection region, \circled{3} determination of the composition of an ion chain, and \circled{4} selectively discarding one of the sub-chains (left/right).
These operations are implemented by applying different sets of voltages to the individual dc electrodes, realizing the required potential landscapes.
The applied voltages in our realization of this method can be seen in \cref{flow-chart}.

\begin{figure}
\includegraphics[width=8.6cm]{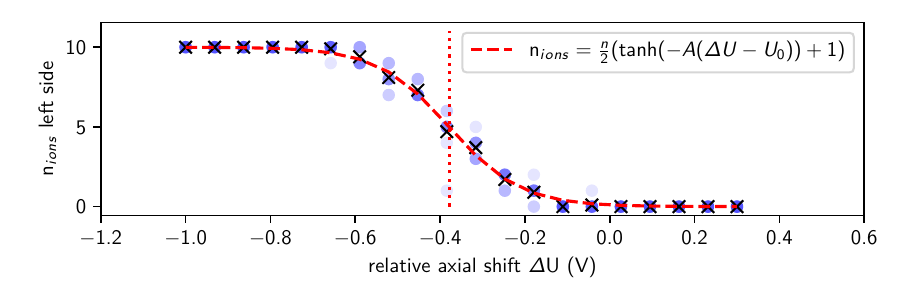}
\caption{
Calibration of the splitting process.
A ten ion crystal was split and the ions on the left side were counted. The measurement was repeated ten times for each value of $\Delta U$. The crosses denote the average ion number detected on the left side. The circles show each measurement result, with its frequency encoded in the visibility of the circle.
The red dashed line shows a fit to the data, which is used to infer the voltage $U_0$, needed to split the crystal into equal parts, shown by the dotted line.
}
\label{fig-ion-chain-calib}
\end{figure}

For the splitting operations, the ions are initially confined in a shallow potential ($\omega_z = 2\pi \times$\SI{143}{\kHz}) and the voltage on the center electrode (V$_3$) is gradually increased, forming a double-well potential that splits the ion chain in two parts. Nomenclature of the blade segments can be found in \cref{trap-cad-nomenclature}.
The ratio of ions in the left and right well of the potential can be controlled by shifting the equilibrium position of the ions in the axial direction before splitting. This is done by applying an additional differential voltage $\Delta U$ to the outer electrodes (V$_1$ and V$_5$).   
  A measurement of the dependence of the splitting ratio on the
  differential voltage is shown in \cref{fig-ion-chain-calib}.
  From these measurements, $\Delta U$ can be set such, that the ion crystal is split in half. More details can be found in Appendix \ref{app-ion-splitting-calib}. This splitting ratio voltage is applied to all ion operations.

The minima of the double-well potential formed at the end of the splitting operation are outside of our imaging region. Therefore for determination of the constituents, a shuttling operation is required to move one of the sub-chains back, while keeping both parts separated.
This is achieved by increasing the outer electrodes' voltage on one of the sides: V$_1$ (or V$_5$) and applying a small trapping potential with the inner electrodes (V$_2$ and V$_4$).

When one of the sub-chains is shuttled into the imaging region, the other sub-chain is trapped in the region of the outer electrode. There, confinement in axial direction is partially provided by the voltage on electrode V$_4$ (or V$_2$) and the axial component of the rf potential in the outer region of the trap.

The last required operation is the ejection of
the unwanted sub-chain from the trap.
This operation is implemented by setting one side of the outer electrodes to \SI{19}{V} at V$_1$ (or V$_5$) and increasing the center voltage, V$_3$, to push the ions, trapped in one of the double-well minima, out of the trap, while the other ion sub-chain remains trapped.

\section{Automation}\label{automation}
Combining the previously discussed operations with the ion detection protocol allows to automate the process for dual-species two-ion preparation. Our implementation of this automation was integrated into the Advanced Real-Time Infrastructure for Quantum physics (ARTIQ) framework \cite{sebastien_bourdeauducq_2022_6619071}.

\subsection{Ion preparation scheme}
The automated two-ion crystal preparation starts by loading a dual-species ion chain.
We select the loading parameters such, that the number of bright ions is larger than the number of dark ions, i.e.: $n_{\text{bright}} \ge n_{\text{dark}} > 0$. Testing whether this starting condition is met is the first step of the preparation algorithm.
Fulfilling this condition prevents errors in the dark ion detection (see \ref{benchmark}) and results in more efficient cooling of the ions.

After checking this starting condition, the main loop for the preparation algorithm starts. First the ion chain is split into two sub-chains of similar size. Both sub-chains are subsequently shuttled to the center segment to determine their constituents.
Afterwards, one of the sub-chains is ejected, with the specific sub-chain determined based on the logical pathway depicted in \cref{table-decision}.
Then the loop starts again with the remaining ions. The algorithm is visualized in \cref{flow-chart}.

The decision which side to discard is made according to the following logic: First, it is checked if any of the two sub-chains fulfills the "start condition" (see above). If this is not the case, the ion chains are merged and the loop is restarted.
If only one of the two sub-chains fulfills the start condition, this sub-chain is kept, the other one is discarded and the loop is restarted.
If both sub-chains fulfill the start condition, further decision factors are taken into account.
First, the number of dark ions are compared. The sub-chain with fewer dark ions is kept, while the other one is discarded, since fewer dark ions are beneficial for the cooling performance.
If both sub-chains do not differ in this parameter, the one with more bright ions is kept. If both sub-chains are identical, the left sub-chain is discarded, with the selection being arbitrary.
This decision process is summarized in \cref{table-decision}.

\begin{table}
\begin{tabular}{|l|c|c|l|}
\hline
Step & \multicolumn{2}{|c|}{condition} & action \\\hline
1 & \multicolumn{2}{|c|}{ $n_\text{bright} \ge n_\text{dark} > 0$} & \\
  & \hspace{0.6cm} left \hspace{0.cm} & right & \\\cline{2-4}
  & False & False & error \\
  & False & True & discard left \\
  & True & False & discard right \\
       & True & True & continue \\ \hline
     2 & \multicolumn{2}{|c|}{$n_\text{dark}^\text{left}>n_\text{dark}^\text{right}$} & discard left \\
       & \multicolumn{2}{|c|}{$n_\text{dark}^\text{left}<n_\text{dark}^\text{right}$} & discard right \\
       & \multicolumn{2}{|c|}{$n_\text{dark}^\text{left}=n_\text{dark}^\text{right}$} & continue \\ \hline
        3 & \multicolumn{2}{|c|}{$n_\text{bright}^\text{left}>n_\text{bright}^\text{right}$} & discard right \\
          & \multicolumn{2}{|c|}{$n_\text{bright}^\text{left}<n_\text{bright}^\text{right}$} & discard left \\
          & \multicolumn{2}{|c|}{$n_\text{bright}^\text{left}=n_\text{bright}^\text{right}$} & discard any side; \\
  & \multicolumn{2}{|c|}{} & we discard left \\ \hline
\end{tabular}
\caption{Decision process for discarding one side of the ion chain after split.}
\label{table-decision}
\end{table}

This loop is repeated until a two-ion crystal with exactly one dark and one bright ion is left.

In addition, we have implemented a verification if the detection worked as expected. If the number of detected bright and dark ions before splitting does not match the number of bright and dark ions detected after the split, the sub-chains are re-merged and the loop is restarted. If the starting condition is no longer fulfilled at the beginning of the loop, the algorithm will stop and raise an error.

\subsection{Experimental results}
The algorithm described above was run, starting from 100 ion chains containing 10 to 20 ions.
Each initial ion configuration was successfully reduced to a dual-species two-ion ion chain.

\begin{figure}
\includegraphics[width=8.6cm]{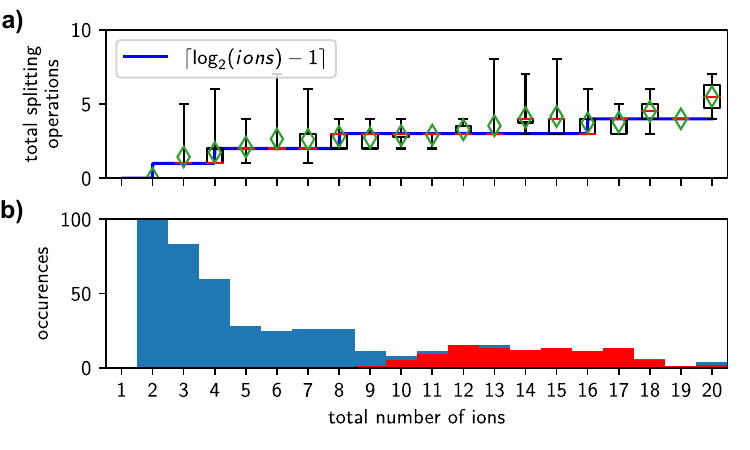}
\caption{
Performance of the two-ion crystal distillation.\\
\textbf{a}$)$ shows the number of splitting operations needed to reduce a given ion crystal to a two-ion crystal. Visualization done via a box-plot due to the asymmetric error distribution.
The whiskers show the min and max value, while the box extends from the first quartile to the third quartile.
The red lines indicates the median and the green diamonds mark the average of the distribution.
The blue curve shows the expected scaling of the algorithm.\\
\textbf{b}$)$ Histogram of the ion number occurrences during the data taking. The histogram gives the sample size for each data point in the plot above.
100 dual-species ion crystal chains were prepared with ion numbers between 10 and 20 (red bars) and reduced to a two-ion crystal using the described algorithm.
The blue bars show the intermediate ion crystal sizes.
}
\label{ion-crystal-alg}
\end{figure}

The reported samples showed no instances where the algorithm failed. It typically took around four splitting operations to arrive at the final ion chain. This translates to a time of $\sim$\SI{100}{\s} in our setup. The runtime is mainly determined by the conservative waiting time of \SI{18}{\s} per splitting cycle to make sure that the ion chain is crystallized.
\cref{ion-crystal-alg} shows the number of split operations needed to arrive at a two-ion crystal, dependent on the size of the initial ion crystal configuration.
During the procedure of forming a two-ion crystal, the algorithm naturally produces ion chains which are smaller than the starting size. These intermediate ion chains were all tracked and interpreted as start configurations. Thereby smaller starting ion chain configurations could be analyzed without producing them by ion loading. With this procedure we could generate around 300 data points from 100 prepared ion crystals.

For a process that ideally splits the ion chain in the center, one would expect the number of splitting operations to follow $\lceil\log_2{(n)}-1\rceil$, with $n$ being the total number of ions. Since only integer values of splitting operations are possible, the ceiling function ($\lceil \ \rceil$) is needed. This estimate is indicated by the blue line in \cref{ion-crystal-alg}. Most median values follow that expected curve.

\section{Outlook}

The presented scheme of ion crystal preparation can be extended to produce ion crystals with an arbitrary dual-species ion composition, 
as long as the compositions of the sub-chains can be determined. This could be achieved by changing the decision tree or by changing the splitting ratio dynamically. Before each split operation, an optimal splitting ratio could be determined and used, in order to prepare the desired ion crystal composition, possibly combined with ion reordering \cite{splatt_deterministic_2009}. A dynamic splitting ratio would also increase the speed of the algorithm.

In the work presented here, only ion crystals up to 20 ions were investigated. Extending the procedure to larger numbers of ions is possible, as the algorithm runtime scales very favorably (logarithmic) with ion number.
For larger ion chains the axial trapping potential would need to be lowered or the radial confinement increased. Alternatively, a more involved detection scheme could be implemented, capable of calculating and detecting ion positions in a 3D crystal \cite{schuessler_3Dioncrystal_2010}.

\section{Conclusion}
We have presented a simple, deterministic and efficient technique for preparing a dual-species, two-ion crystal from a large dual-species ion crystal and demonstrated an automated implementation.
The method at hand produces a two-ion crystal in less than two minutes. It alleviates the need for fine tuning of loading parameters, which increases the robustness and reproducability of preparing dual-species two-ion crystals. Our method is easily extendable to produce different configurations of dual-species ion crystals and can be further optimised for speed.
The presented scheme relies on experimental techniques that are available in Quantum Charge Coupled Device (QCCD) architectures for quantum computing \cite{kielpinski_architecture_2002, pino_demonstration_2020} and are well suited for trapped ion quantum technology due to the demonstrated high level of automation.

\section{Acknowledgments}
We want to thank Jan Kiethe and Elizaveta Surzhikova for developing and providing the readout script for the EMCCD camera, which was used in this work.
We acknowledge the additional dark ion detection insights from Leon Schomburg and Tabea Nordmann.
Additionally, we thank Lukas Spie^^c3^^9f for careful reading of the manuscript and for providing valuable suggestions and Kai Dietze for helpful support with the Artiq experimental control.
This research was funded by the Deutsche Forschungsgemeinschaft (DFG, German Research Foundation) ^^e2^^80^^93 Project-ID 274200144 ^^e2^^80^^93 SFB 1227 (DQ-mat), project B05 and B03 with partial support from Germany's Excellence Strategy EXC-2123 QuantumFrontiers 390837967. This project has received funding from the European Research Council (ERC) under the European Union^^e2^^80^^99s Horizon 2020 research and innovation program (grant agreement No 101019987). This work was partially funded by 22IEM01 TOCK. The project (22IEM01 TOCK) has received funding from the European Partnership on Metrology, co-financed from the European Union^^e2^^80^^99s Horizon Europe Research and Innovation Programme and by the Participating States.

\section{Author Contributions}
    T. R., M. J. Z. and F. W. conceived the main idea of this research and carried out the experiments. Furthermore T. R., M. J. Z., P. O. S. and F. W. advanced the work by interpreting the results, steering the research and writing the manuscript. J. K. and T. E. M. contributed the idea of the dark ion detection principle and proved its usefulness in another experimental setup.
    T. R. and M. J. Z. contributed equally to this work.

\bibliographystyle{apsrev4-2}

\bibliography{all-3}

\appendix
\section{Transformation from calculated ion positions to camera pixels}\label{sec:app-cam-calib}
The theoretically determined position of the $j$-th ion with respect to the center of the ion chain is given by $x'_j=lu_j$, where $u_j$ is the dimensionless equilibrium position that can be computed analytically for two and three ions and needs to be calculated numerically for larger ion chains \cite{james_quantum_1998}. The length scale $l$ is for singly ionized ions given by
\begin{align}
l^3=\frac{e^2}{4\pi\epsilon_0 C_\mathrm{trap}U_\mathrm{dc}},
\end{align}
where $e$ is the charge of the electron, $\epsilon_0$ is the permitivity of free space, $C_\mathrm{trap}$ is a constant given by the geometry of the ion trap and $U_\mathrm{dc}$ is the voltage that provides the axial confinement.

The position of the image on the camera $P^\text{theo}_j$ is related to the ion's position $u_j$ by $P^\text{theo}_j = P_0+Ku_j$, where $P_0$ denotes the center of the ion chain on the camera chip and $K=lM/s_\mathrm{px}$ is a factor that accounts for the magnification factor $M$ of the imaging system and the pixel size $s_\mathrm{px}$. The values for $P_0$ and $K$ are inferred from loading a known number of fluorescing ions ($\ge 2$) and determining their position on the camera image.

\section{Split ratio calibration}\label{app-ion-splitting-calib}
Measuring the amount of ions on one side of the double-well potential for different differential voltages $\Delta U$ between electrodes V$_1$ and V$_5$ and fitting a model to the data, allows to find a value for $\Delta U$ matching the desired splitting ratio. Here we are interested in a splitting ratio of 1.

The amount of ions in the left well after the split, resembles the heuristic function
\begin{align}
\label{fit_function}
n_\text{left}(\Delta U) = \frac{n}{2}\tanh\left( -A \left( \Delta U-\Delta U_0 \right)+1\right),
\end{align}
where $n$ is the total number of ions, $A$ a scaling factor and $\Delta U_0$ is the differential voltage, for which the ion crystal is split in half. A fit of \ref{fit_function} to measured data can be found in \cref{fig-ion-chain-calib}.

\section{Detection errors}\label{app-detection-errors}
\begin{figure*}
\includegraphics[width=0.4\textwidth]{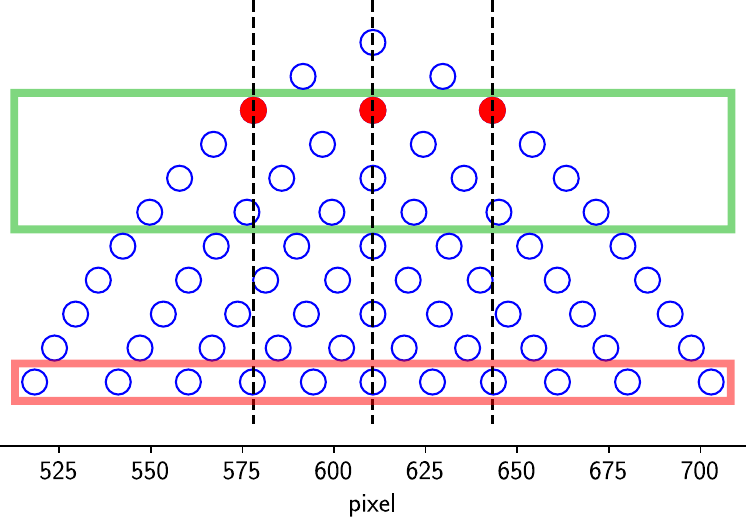}
\caption{
Example of configurations that are difficult to distinguish. Shown are calculated ion positions for ion chains with ion numbers between 1 and 11. Circles illustrate ion positions. Assuming three trapped bright ions (red circles), there is a configuration with 11 ions, which includes 3 ion positions almost matching the positions of the 3 trapped ions. The green rectangle marks the region of interest of our dark ion detection scheme, applying the limiting ratio of $n'_\mathrm{dark}/n'_\mathrm{bright}=1$. The black lines mark the position of the three bright ions.
Without the limiting ratio, the dark ion detection scheme can not reliably distinguish between the 3 bright ion configuration and the 11 ion configuration marked in red with 3 bright located at the same locations as the ions in the 3 ion crystal and 8 dark ions.
}
\label{detection-error-example}
\end{figure*}

When detecting the ion chain configuration, errors can occur if there is no limit to the amount of dark ions.
Most of these errors originate in ion chain configurations for which some ion positions are similar for different configurations. In this context, similar means that the position cannot be differentiated with the given experimental position resolution. If only these similar ion positions are occupied by bright ions, the different configurations are indistinguishable to the algorithm.
For large ion crystals the density of ions increases and thereby those ion chains often have a subset of ion positions which are almost identical to some ion positions of a smaller ion crystal.
The most obvious example of a position which is similar in many ion chains, is the center position in an odd-numbered ion crystal.
If only one fluorescing ion is trapped, it is impossible to determine the total chain size, since non fluorescing ions could be present and located on each side of the bright central ion.
An illustration for a typical false detection can be found in \cref{detection-error-example}.

These errors can be mitigated by limiting the allowed ion chain configurations to ones with $n'_\mathrm{bright} \ge n'_\mathrm{dark}$ or by a more accurate bright ion position determination (this would however not solve the problem with only one fluorescing ion in the center).

\end{document}